\documentclass[11pt]{article}
\usepackage[latin1]{inputenc}
\usepackage{epsfig}
\usepackage{amsmath,amsfonts,latexsym, amssymb}

\newtheorem{satz}{Theorem}[section]

\newtheorem{conclusion}[satz]{Conclusion}
\newtheorem{ob}[satz]{Observation}

\newtheorem{statement}[satz]{Statement}

\newcommand{\mbf}{\mathbf}

\newcommand{\tit}{\textit}

\begin{document}
\thispagestyle{empty}
\begin{center}
\vspace*{1.0cm}

{\LARGE{\bf Thermodynamics meets Special Relativity\\
-- or what is real in Physics?}} 

\vskip 1.5cm

{\large {\bf Manfred Requardt }} 

\vskip 0.5 cm 

Institut f\"ur Theoretische Physik \\ 
Universit\"at G\"ottingen \\ 
Friedrich-Hund-Platz 1 \\ 
37077 G\"ottingen \quad Germany\\
(E-mail: requardt@theorie.physik.uni-goettingen.de)

\end{center}

\vspace{0.5 cm}

\begin{abstract}
  In this paper we carefully reexamine the various framworks existing
  in the field of relativistic thermodynamics. We scrutinize in
  particular the different conceptual foundations of notions like the
  relativistic work, heat force, moving heat and relativistic
  temperature. As to the latter notion we argue that, as in ordinary
  thermodynamics, relativistic absolute temperature should be
  introduced operationally via relativistic Carnot processes. We
  exhibit the more implicit or even hidden tacit preassumptions being
  made and point to a couple of gaps, errors and inconclusive
  statements in some of the existing literature.  We show in
  particular that there is a wide-spread habit to draw general
  conclusions from the analysis of too restricted and special
  thermodynamic processes, e.g.  processes with constant pressure,
  which is dangerous and sometimes leads to wrong results.
  Furthermore, we give a detailed analysis of the so-called zeroth law
  of relativistic thermodynamics with the help of a relativistic
  Carnot process. We rigorously show that, contrary to certain
  statements in the literature, thermodynamic systems at different
  relativistic temperatures, moving relative to each other, can
  thermally stably coexist provided that their respective temperatures
  obey a certain functional relation (given by the Lorentz factor).
  This implies however that their respective rest temperatures are the
  same.

\end{abstract} \newpage
\setcounter{page}{1}
\section{Introduction}
While relativistic thermodynamics is in principle a quite old field of
research, starting almost immediately after the fundamental Einstein
papers of 1905 with seminal contributions by Planck, Einstein and
Planks student v.Mosengeil (see for example
\cite{Planck1},\cite{Einstein},\cite{Mosengeil}), there is
nevertheless a still ongoing debate both about the overall working
philosophy, certain of its basic principles and various technical
details. See for example the recent \cite{Haenggi}, which, however,
deals primarily with various problems of relativistic statistical
mechanics (a catchword being: Juettner distribution). In the following
we will refrain from commenting on the many additional problems being
inherent in the latter field, as relativistic thermodynamics is
already a quite ambitious field of its own. Furthermore, while there
exist of course a lot of connections between thermodynnamics and
statistical mechanics, the relativistic regime poses quite a few
problems of its own due to the relatively rigid constraints on the
class of admissible microscopic interactions if one stays within the
framework of (point) particles. Conceptually it may therefore be
reasonable to regard relativistic quantum field theory as the
appropriate framework to develop a relativistic version of statistical
mechanics with its natural possibilities of particle creation and
annihilation and the interaction of fields replacing the forces
between (point) particles.

The reasons for this still ongoing debate in relativistic
thermodynamics are manifold. The history of the different points of
view and approaches is in our opinion meanwhile so contorted and
facetted because two fields had to be merged which have their own
specific technical and epistomological problems. It hence appears to
be reasonable to us, to try to isolate the crucial points where
opinions differ and concentrate on the deeper reasons, why discussions
have lasted for such a long time without coming to a final conclusion.
This holds in particular so as we will show that in our view various
of the common arguments do contain gaps and even errors, which we try
to exhibit in the following.

Our own interest was raised anew when we came across the so-called
``Einstein-Laue Discussion'' as being reviewed in \cite{Treder} and
\cite{Liu}. It is a curious but little known fact that according to
the detailed analysis of the exchange of letters between Einstein and
Laue, made by Liu, Einstein changed his opinion about the correct
transformation properties of various thermodynamic quantities
completely in the early fifties without apparently being aware of this
fact. While in \cite{Einstein} he got results which go conform with
the results of e.g. Planck, he arrived already in 1952 at
transformation laws which a couple of years later were published by
Ott and Arzeli\`{e}s (\cite{Ott},\cite{Arzelies1}).

While this may be a remarkable psychological or historical phenomenon,
what is conceptually more important in our view, is the deeper reason
why such eminent thinkers came to contradictory conclusions, as it can
certainly be ruled out that for example Planck, Einstein, Laue, Pauli,
Tolman, to mention a few, simply commited errors in their
calculations. It therefore seems to be worthwhile, to analyse the
steps in the reasonings of the various authors who contributed to this
field and to exhibit and isolate the sometimes only tacitly made or
even entirely hidden preassumptions on which the various analyses were
based. It then becomes perhaps clearer, in what sense our subtitle: ``or
what is real in physics'',  may be justified.

In our view one of the problems in the more recent discussions is in
fact that the respective physical situation is frequently only
incompletely described, or, on the other hand, a very special case is
analysed instead of a really general move on the thermodynamic state
manifold, thus leaving out important aspects or emphasizing only
points which support the own point of view. We will come to this
phenomenon in more detail in the following sections but mention just a
typical example, i.e. the controversial discussion between
Arzeli\`{e}s, Gamba and Kibble
(\cite{Arzelies1},\cite{Gamba},\cite{Kibble} and the respective
comments and remarks in the same volume of the journal). When reading
for example these papers, it becomes obvious that the authors simply
talk about quite different systems and incompatible situations, while
apparently being only incompletely aware of this fact. Anyhow, in our
opinion the position of Kibble is the more reasonable one in the
mentioned discussion.

To give an example, we think it is not helpful to actually include
parts of the exterior of the confined system or the walls into the
thermodynamic discussion. One should rather adopt the philosophy that
thermodynamic systems are dealt with in the way they are defined in
ordinary thermodynamics. Otherwise the discussion becomes very
cumbersome in our view. In this context we would like to remark that
our criticism applies also to certain points in the paper of Ott. We
will comment on these aspects in section \ref{Ott}.

To begin with, we make a brief classification of the different working
philosophies and opinions (see also \cite{Haar}). First, there is the
\tit{classical period}, represented e.g. by Planck, Einstein, v.Laue,
Pauli, Tolman
(\cite{Planck1},\cite{Einstein},\cite{Laue},\cite{Pauli},\cite{Tolman}),
and being roughly described by the transformation laws of heat and
temperature
\begin{equation}\delta Q= \delta Q_0/\gamma\quad , \quad T=T_0/\gamma
\end{equation}
with $\gamma$ the \tit{Lorentz factor}
\begin{equation}\gamma=\left(1-u^2/c^2\right)^{-1/2}        \end{equation}
and the subscript, $0$, denoting in the following the variables in the
comoving inertial or rest frame (CIF) of the thermodynamic system. Its
velocity relative to the laboratory frame is $u$ and the variables in
the laboratory frame are $\delta Q$ and $T$.

For convenience we usually assume that the thermodynamic system is at
rest in the IF, $X'$, which is in \tit{standard position} with respect
to the laboratory frame, $X$. That is, it moves with velocity $u$ in
the positive x-direction, its coordinate axes being parallel to the
ones of $X$ and with coordinate origins coinciding at $t=t'=0$
(\cite{Rindler}).

In the sixties (and earlier in the mentioned letters of Einstein)
another transformation law emerged (\cite{Ott},\cite{Arzelies1}):
\begin{equation}\delta Q= \delta Q_0\cdot\gamma\quad , \quad T=T_0\cdot\gamma       \end{equation}
One should however remark that, while superficially being the same,
the point of view of e.g. Arzeli\`{e}s is quite different from the
one, hold by Ott (cf. the discussion between Arzeli\`{e}s, Gamba and
Kibble, mentioned above). Furthermore, while we arrive at the same
transformation laws in the present paper as Ott, the situation
discussed by him in \cite{Ott}, section 2, is also only a particular
case and does not! really deal with a general variation of
thermodynamic variables. Therefore some of the really critical
problems were not
addressed by him (cf. the section about the Ott-paper). \\[0.3cm]
Remark: Note that Moeller in his beautiful book (\cite{Moeller})
changed his convention from the classical point of view in the earlier
editions to the convention of Ott and Arzeli\'{e}s in the last edition
which we are citing in the references.\vspace{0.3cm}

Somewhat later and up to quite recent times a third approach was
promoted by e.g. Landsberg and coworkers
(\cite{Landsberg1},\cite{Landsberg2},\cite{Landsberg3},\cite{Landsberg4},\cite{Matsas};
see also \cite{Gron}), another reference, discussing various points
of view is \cite{Kampen}. Landsberg et al argue that temperature and
heat are \tit{Lorentz-invariants}, i.e. behave as scalars, that is
\begin{equation}\delta Q= \delta Q_0\quad , \quad T=T_0
\end{equation}
This point of view is presently shared by a number of other workers in
the field. It is sometimes argued that all this is rather a matter of
convenience as the transformation laws are not really fixed by the
condition of relativistic covariance. This is certainly correct. On
the other hand, some of the arguments advanced in favor of this latter
opinion are in our view not really convincing. As a prominent
reference for such an opinion (``pseudoproblem'') see for example
\cite{Sexl}, p.334.

It is e.g. argued that both the classical point of view
($T=T_0\cdot\gamma^{-1}$) and the more modern one ($T=T_0\cdot\gamma$)
are compatible in so far as in the former heat is exchanged at
constant velocity, in the latter at constant momentum. We will show in
the following that this parallelism of points of view cannot be
maintained. We rigorously show that the heat force as it was invoked
in the classical approach does simply not exist (section \ref{work}),
more specifically, of the three components making up the total
classical heat force, only one can be granted a real existence.
Furthermore, exchange of heat between system and comoving reservoir
(i.e., both having the same velocity) is a transparent process. What
however is the meaning of exchange of heat at equal or constant
momentum between a (possibly) small system and a huge reservoir? We
show below that we get the transformation laws
\begin{equation}\delta Q=\delta Q_0\cdot\gamma \quad , \quad
  T=T_0\cdot\gamma    \end{equation}
via heat exchange at constant velocity.  

What, for example, frequently happens is that two, in principle
different, situations are mixed up. If one inserts an empirical
thermometer into a moving substance and reads off the temperature from
the laboratory frame, there exists little doubt that one in fact
observes the \tit{rest temperature} $T_0$. This is not! the
temperature of a moving system. In this respect temperature behaves
differently from length or time. Another thought experiment
(\cite{Landsberg3} and elsewhere) discusses the heat exchange between
bodies moving relative to each other and tries to construct a
paradox unless the temperature is a scalar. This argument is also
flawed as we will show in section \ref{zeroth} (the \tit{relativistic zeroth
  law}).

At the end of this introduction we want to briefly comment on two
other papers. In \cite{Horwitz} it is for example argued that one
should take the thermodynamical variables as scalars \tit{like in
  general relativity}? In the first place, the building blocks of
general relativity are general \tit{tensors}. Scalars do not play any
particular role. In our approach some of the variables are 4-vectors
which have a very nice transformation behavior. Furthermore, it is
claimed that some \tit{grotesque situations} do arise because of the
non-equivalence of simultaneity. As to this point, it is frequently
overlooked that by assumption all moves on the state manifold are
performed in a quasi-static way so that the problem of
non-simultaneity is not really virulent. 

In \cite{Eimerl} the author introduces a new principle, claiming that
thermal equilibrium between bodies in relative motion is
impossible. We show however in the section about the zeroth law of
thermodynamics (section \ref{zeroth}) that, to the contrary, this is possible
and free of logical contradictions if described in the way we did it
and as it was anticipated by v.Laue. 
\section{Notations and Standard Formulas in Relativistic Mechanics}
The Minkowski metric is denoted by $\eta_{\nu\mu}$ with the convention
$(+ - - -)$. Greek indices
 run from $0$ to $3$, latin indices from $1$
to $3$. Four-vectors carry a greek index (abstract index notation),
three-vectors are in bold face. The energy of a particle or system is
denoted by $E$, its three-momentum by $\mbf{G}$ (as the letter $p$ is
already used for the pressure). We write
\begin{equation}ds^2=c^2\cdot d\tau^2=\eta_{\nu\mu}dx^{\nu}dx^{\mu}      \end{equation}
with $x^0=c\cdot t$ and $d\tau$ the proper time interval measured by a
comoving ideal standard clock (i.e., a clock being unaffected by
acceleration; see the discussion in e.g. \cite{Rindler}).

We have 
\begin{equation}E=m\cdot c^2\quad , \quad \mbf{G}=m\cdot \mbf{u}\quad
  ,\quad m=\gamma\cdot m_0=m_0\cdot (1-u^2/c^2)^{-1/2}       \end{equation}
for a particle or system moving with the momentary velocity $\mbf{u}$
relative to a certain IF. $m_0$ is the proper mass and $\gamma$ the
\tit{Lorentz factor}. In 4-vector notation this reads
\begin{equation}G^{\nu}=(E/c,\mbf{G})=m_0\cdot U^{\nu}      \end{equation}
with 
\begin{equation}U^{\nu}=dx^{\nu}/d\tau=(c\cdot\gamma,\mbf{u}\cdot\gamma)     \end{equation}
the 4-velocity with $\eta_{\nu\mu}U^{\nu}U^{\mu}=c^2$.

The 3-force is conventionally defined via
\begin{equation}\mbf{F}=d/\!dt\,(m\cdot u)=d/\!dt\,\mbf{G}      \end{equation}
Its transformation properties (and the covariance properties of other
derived notions) become more transparent by finding the correct
4-dimensional generalisation. The 4-force (or Minkowski force,
\cite{Minkowski}) is defined by
\begin{equation}F^{\nu}=(\gamma\cdot\mbf{F}\cdot\mbf{u}/c,\gamma\cdot\mbf{F})       \end{equation}
with $F^{\nu}=dG^{\nu}/d\tau$. Note that it holds (for a
\tit{rest-mass preserving force}!)
\begin{equation}\mbf{F}\cdot\mbf{u}=\mbf{F}\cdot d\mbf{x}/dt=dW/dt=dE/dt       \end{equation}
so that in that case the 4-force can alternatively be written
\begin{equation}F^{\nu}=(\gamma\cdot c^{-1}\cdot dE/dt,\gamma\cdot\mbf{F})     \end{equation}
This subtle point is treated in more detail in subsection \ref{work}
where we discuss the relativistic concept of work in more detail. Note
that in collisions the rest-mass may change in the moment of
contact. Rindler in \cite{Rindler2}, p.92 speaks in this context of
\tit{heat-like} forces.

As in relativistic mechanics 4-momentum is conserved for a closed
system of particles
\begin{equation}\sum_i m_i\cdot \mbf{u}_i= const      \end{equation}
implies that the above definition of force guarantees that the
law:actio $=$ reactio holds. This plays a certain motivational role for
conceptual generalisations being made in relativistic thermodynamics
and is particularly stressed by Tolman (\cite{Tolman}).

Defining acceleration or 4-acceleration by
\begin{equation}\mbf{a}=d\mbf{u}/dt\quad , \quad
  a^{\nu}=dU^{\nu}/d\tau=\gamma\cdot dU^{\nu}/dt      \end{equation}
one sees, that in general the force vector is not! parallel to the
acceleration vector. We rather have
\begin{equation}\mbf{F}=m\cdot d\mbf{u}/dt+dm/dt\cdot\mbf{u}     \end{equation}
The (problematic) generalisation to relativistic thermodynamics will
also play a certain role in the following.
\section{Some Formulas from Relativistic Continuum Mechanics}
The derivation of the thermodynamic behavior of relativistic
equilibrium systems is to a large extent based on concepts from
continuum mechanics. A very well-written source are the chapters 6 and
7 in \cite{Moeller}, which we recommend as a reference. As already at
this stage some diverging opinions emerge, being related to various at
first glance counterintuitive aspects of the theory (cf. e.g. the above
mentioned dispute between Kibble, Arzeli\`{e}s and Gamba ), some brief
remarks seem to be in order.
 
Treating everything within the well developed framework of field
theory has a great advantage. There exists a rich source of notions
and calculational tools which are founded on well understood
principles. The most important concept is the
\tit{energy-momentum-stress tensor}, $T_{\nu\mu}$ , which is the
starting point of most of the derivations. For closed systems (cf.
\cite{Moeller} chapt. 6) it has the important property that it leads
to conservation of energy-momentum and that its symmetry leads to the
canonical identification of momentum density and energy-flow.

Furthermore, for \tit{stressed} continuum systems it allows for the
transparent derivation of somewhat counterintuitive formulas. For the
momentum density we have 
\begin{equation}\partial_t g_i+\partial_{x_k}(g_i\cdot u_k+t_{ik})=0   \end{equation}
with
\begin{equation}T_{ik}=(g_i\cdot u_k+t_{ik})       \end{equation}
Here $\mbf{g}$ is the momentum density, $\mbf{u}$ the local velocity
(relative to some IF) and $t_{ik}$ the \tit{(relative) stress
  tensor}. Note that the occurrence of this latter term is at first
glance perhaps a little bit unexpected, that is, stress density
contributing to momentum flux. Correspondingly we have for the energy
flux:
\begin{equation}\partial_t \varepsilon+\partial_{x_k}(\varepsilon\cdot
  u_k+u_i\cdot t_{ik})=0        \end{equation}
with 
\begin{equation}S_k:=(\varepsilon\cdot
  u_k+u_i\cdot t_{ik})      \end{equation}
the energy flux.\\[0.3cm]
Remark: Note that the energy density includes the elastic contributions in
addition to the translatory energy of ordinary moving
matter.\\[0.3cm]
One should make a remark as to the two versions of stress tensor. The
above version is called the \tit{relative stress tensor}. It has a
transformation behavior which is different! from the covariance
behavior of the so-called \tit{absolute stress tensor},
\begin{equation}p_{ik}=g_i\cdot u_k+t_{ik} \end{equation} The letter
version occurs in the full energy momentum tensor of field theory and
really has the correct transformation behavior of a 2-tensor under the
Lorentz group (cf. the remarks in \cite{Moeller}, p.184ff or
\cite{Tolman}, p.69ff). In $t_{ik}$ forces or stresses are calculated
relative to a surface element being momentarily at rest with respect
to the medium!, while in the latter case a coordinate system is used
which is e.g. at rest in space (Lagrange versus Euler point of view in
continuum mechanics).

According to the general principles of relativistic (system)
mechanics, we have the canonical identification 
\begin{equation}g_k=S_k/c^2      \end{equation}
Remark: Given the transparent and straigthforward derivation of these
formulas (as far as we know, being due to v. Laue), it is a little bit
surprising that e.g. the stress contributions in the energy flux are
called \tit{pseudo contributions} by Arzeli\`{e}s.\vspace{0.3cm}

This point becomes particularly important if a thermodynamic
equilibrium system, being enclosed in a container, is treated with the
stresses being reduced to a scalar pressure, $p$. In that case the
expression for the momentum density becomes
\begin{equation}g_k=\rho\cdot u_k+p\cdot u_k/c^2        \end{equation}
with $\rho$ the relativistic matter density, $\varepsilon=\rho\cdot c^2$.

In the particular case of thermodynamic equilibrium systems, the
pressure is constant over the volume of the system. Furthermore, we
assume that it moves uniformly with velocity $\mbf{u}$. We then can
easily integrate the above equations and get:
\begin{equation}\mbf{G}=(E+p\cdot V)/c^2\cdot \mbf{u}      \end{equation}
What is not yet known is the functional form of the energy, $E$.

In \cite{Tolman} $\varepsilon$ or $E$ is calculated in the following
way. With the definition of 3-force, $\mbf{F}:=d\mbf{G}/dt$, we have
\begin{equation}dE'/dt=\mbf{F}'\cdot\mbf{u}'-\cdot dV'/dt
\end{equation} for intermediate values of the respective variables.
We start from a system which is initially at rest, having pressure
$p^0$ and proper volume $V^0$ and which is then quasi-stationary
brought to the final velocity $\mbf{u}$. Using the above expression
for $\mbf{G}$, the fact that $p$ is a Lorentz invariant, i.e. $p=p_0$,
and the change of volume by Lorentz contraction, $V=V_0\cdot \gamma$,
one can integrate the above expression and get
\begin{equation}E+p\cdot V=(E_0+p_0\cdot V_0)\cdot\gamma      \end{equation}
or
\begin{equation}\label{25}E=(E_0+p_0V_0\cdot u^2/c^2)\cdot\gamma      \end{equation}
as the desired transformation equation for energy.

Another, in our view more direct, method goes as follows. Starting
again from the rest system, the work done by pressure forces due to
Lorentz contraction is ($p=p_0$):
\begin{equation}\label{27}\Delta E_1=-p\cdot (V-V_0)=p_0\cdot V_0\cdot
  (1-\gamma^{-1})=
(p_0V_0\cdot u^2/c^2)\cdot\gamma     \end{equation}
The translatory contribution is 
\begin{equation}E_2=E_0\cdot\gamma     \end{equation}
hence
\begin{equation}\label{28}E= (E_0+p_0V_0\cdot
  u^2/c^2)\cdot\gamma\quad\text{or}\quad E+pV=(E_0+p_0V_0)\cdot\gamma    \end{equation}
Remark: In the latter derivation we have not used the definition of
force. On the other hand, concerning the question: what is real in
physics?, we have calculated the work done by the pressure on the
volume, changing due to Lorentz contraction. Hence, people who
consider this as being only apparent (e.g. Rohrlich,\cite{Rohrlich}), may have problems with the above
derivations.
\begin{ob}\label{enthalpy}We see that for non-closed systems like an equilibrium
  system, being confined to a vessel, $(E/c,\mbf{G})$ does not
  transform like a 4-vector, as might be expected from ordinary
  relativistic kinematics. On the other hand, $(H/c,\mbf{G})$ is a
  4-vector, with
\begin{equation}H:=E+pV    \end{equation}
the \tit{enthalpie}. 
\end{ob}
The reason is that work is done by e.g. the walls of the vessel or the
exterior, which is not included in $E$. The other possibility is to
include all! contributions in the system under discussion, for example
the contributions of the walls, and use a total energy-momentum tensor
as in the case of a closed system. This would however become a very
nasty enterprise in our view and should be avoided.
\section{The First and Second Law of Relativistic Thermodynamics}
In this section we describe on what fundamental laws we want to base
relativistic thermodynamics. We begin with the first and second law,
because the zeroth law is more complicated to formulate. It is useful
to divide the set of mechanical or thermodynamical variables into
\tit{invariants} and \tit{covariants}, respectively. We regard entropy
and pressure as invariants under Lorentz transformations.
\begin{equation}S=S_0\quad , \quad p=p_0 \end{equation} This can be
directly calculated for the pressure via its definition as force per
area. In case of the entropy, $S$, one can also provide a
calculational argument (as already Planck did). We prefer however to
invoke the statistical nature of entropy as a measure of the width of
the distribution relevant microstates, being represented by a
particular macrostate.  This property will not change if systems are
quasi stationarily accelerated while maintaining their interior state
(see also \cite{Laue} chapts.4.e and 23).\\[0.3cm]
Remark: As to the definition of pressure as an invariant, the
following subtle point should be kept in mind. This holds for a
concept of pressure being defined with respect to a surface element
being momentarily at rest in the medium! The pressure, occurring e.g.
in the energy-momentum-stress tensor is of course not! a scalar but it
contains an additional kinematical term (cf. e.g. \cite{Moeller}). It
is defined with respect to a coordinate system being fixed in space or
space-time.\vspace{0.3cm}

We assume the first and second law of thermodynamics to hold also for
moving equilibrium systems. For the rest system we have
\begin{equation}dE_0=\delta Q_0+\delta W_ 0=T_0\cdot dS_0-p_0\cdot
  dV_0 \end{equation}
 where for reasons of simplicity we assume all
processes to be reversible. We assume that corresponding laws exist
for the moving system, whereas some of the variables have yet to be
scrutinized in more detail in this latter case. That is,
\begin{equation}dE=\delta Q+\delta W\quad , \quad \delta Q=T\cdot dS    \end{equation}
in the reversible case.

The meaning of $E,p,V$ is clear. More problematical is the meaning of
$\delta Q,T,\delta W$. A very naive first guess can immediately be
ruled out. One may be tempted to assume that all contributions in the
first law do transform in the same way under Lorentz transformation
(as they are of the same nature). Assuming for example that $dE$ is
the zeroth component of a 4-vector (as in ordinary relativistic
mechanics; but see the preceding section), and that $\delta W$ has the
same form as in the rest system, i.e. $\delta W=-p\cdot dV$, we would
get:
\begin{equation}dE=dE_0\cdot\gamma\quad , \quad \delta
  W=-pdV_0/\gamma=\delta W_0/\gamma    \end{equation}
We immediately can infer from this observation that the whole matter
must be more complicated.
\subsection{\label{work}The Relativistic Concept of Work}
In relativistic mechanics the 3-force was canonically defined by
$\mbf{F}=d\mbf{G}/dt$, and this identification was taken over
unchanged and, apparently without much hesitation, by all workers of
the classical period (e.g. Planck, Einstein, v.Laue, Pauli, Tolman) to
the regime of relativistic thermodynamics. This is a little bit funny,
because we will see immediately that it leads to strange consequences,
which were however fully accepted by the above mentioned
scientists. They even found strong arguments why these strange
consequences are in fact entirely natural. 

With
\begin{equation}\mbf{F}=d\mbf{G}/dt=d/\!dt(m_0\cdot\gamma\cdot \mbf{u})=
\dot{m}_0\cdot (\gamma \mbf{u})+m_0\cdot d/\!dt(\gamma\mbf{u})
\end{equation}
there may be a non-vanishing contribution even if the velocity remains
constant, provided $\dot{m}_0\neq 0$. This extra and counter intuitive
term was greeted by e.g. Tolman in chapt. 25 of \cite{Tolman} as a
contribution which may come from a possible influx of heat at constant
velocity.

In adiabatic or purely mechanical processes, where the rest mass or
rather rest energy remains constant, we would have the ordinary
(Newtonian) interpretation of 3-force
\begin{equation}\dot{E}=\dot{W}=\mbf{F}\cdot \mbf{u}      \end{equation}
with $dW=\mbf{F}\cdot d\mbf{r}$ the element of mechanical work. This
comes about as follows:
\begin{multline}\label{36}\dot{W}=d/\!dt(m\mbf{u})\cdot\mbf{u}=m\dot{\mbf{u}}\cdot\mbf{u}+\dot{m}\mbf{u}^2
=m_0\gamma\dot{\mbf{u}}\cdot\mbf{u}+m_0\gamma^3\mbf{u}^2/c^2dot{\mbf{u}}\cdot\mbf{u}\\
=m_0\gamma^3\dot{\mbf{u}}\cdot\mbf{u}\cdot
(1-\mbf{u}^2/c^2+mbf{u}^2/c^2)=m_0\gamma^3\dot{\mbf{u}}\cdot\mbf{u}=d/\!dt(m\cdot
c^2)=dE/dt  \end{multline}

Such a type of force is called by Rindler (\cite{Rindler}, p.124 ff) a
\tit{rest-mass preserving force}. In situations where the rest mass is
allowed to vary, we would get an extra contribution,
$\dot{m}_0\gamma\mbf{u}^2$, which cannot be incorporated in
$dE/dt=d/\!dt(m\cdot c^2)$. So, in this case, the identification of
$\mbf{F}\cdot\mbf{u}$ and $\dot{E}$ does no longer hold. 
In the classical papers this distinction is, as far as we can see, not
made, i.e. the classical work term is simply assumed to be
\begin{equation}\delta W:=\mbf{u}\cdot d\mbf{G}       \end{equation}
as the element of work, done on the system.

In non-relativistic thermodynamics work can be done on the system by
pressure forces, $-pdV$, and by adiabatic, translatory forces. In the
classical framework of relativistic thermodynamics a third form of
work is assumed to occur, i.e. a contribution of the above type,
$\delta W:=\mbf{u}\cdot d\mbf{G}$, which can be different from zero
even if we have only an influx of heat at constant velocity,
$\mbf{u}$. This increases the internal energy and hence, by the
relativistic identification of mass and energy, the rest-mass of the
system and thus the momentum. The corresponding force according to
this philosophy is then
\begin{equation}\mbf{F}_{heat}=\dot{m}_0\cdot (\gamma\cdot \mbf{u})       \end{equation}

It is in our view difficult to understand why the classical authors
emphatically justified the occurrence of such a work term. It is
perhaps noteworthy that Moeller in his first editions of
\cite{Moeller} also followed this line of reasoning, while in his last
edition he changed to the Ott-Arzeli\`{e}s convention. Even more
recently Rindler in \cite{Rindler2}, p.92 explicitly states that an
object, being heated in its rest frame, experiences a force in every
other IF of the type described above. On the other hand, Einstein in
his mentioned letters to v.Laue argued against such \tit{virtual}
force terms and also Ott provides some arguments. However, we think
that our above observation (formula (\ref{36})) is perhaps the most
convincing from a theoretical point of view.
\begin{ob}As only a rest-mass preserving force can be associated with
  a true work term in the ordinary mechanical sense and a fortiori
  with a corresponding energy increase, there is no evidence that
  other forms of energy increase should be associated with a force.
\end{ob}

Some other arguments are as follows:
\begin{itemize}
\item In case some systems had different velocities, heat transfer
  between them, which is ultimately the effect of e.g. a large number
  of random interactions between the respective surface constituents,
  would also involve the tranfer of some net momentum etc. However, in
  case the systems have the same velocity, these random exchanges are
  on average undirected so that no momentum transfer should be
  involved.
\item In our view the phenomenon is not even a relativistic one. If
  the combination of two subsystems of the same velocity does form a
  compound system, one can transfer heat or matter from one subsystem
  to the other one, without any effect on the respective
  velocities. It is difficult to see any force being involved.
\end{itemize}

However, this does not! mean that the term $\mbf{u}d\mbf{G}$ has
always taken to be zero in relativistic thermodynamics. The situation
is in fact more complicated (which was, in our view, not even fully
realised in the detailed analysis by Ott; see below). We have,
according to our previous formulas
\begin{equation}\mbf{u}d\mbf{G}=u^2/c^2\cdot
  d(E+pV)=u^2/c^2\cdot\gamma\cdot (E_0+p_0V_0)      \end{equation}
We have argued above, as also Ott did, that $dE_0$ does not! generate
a contribution in the work, induced by some \tit{heat
  force}. Therefore, the term
\begin{equation}u^2/c^2\cdot\gamma\cdot (dp_0V_0+p_0dV_0)       \end{equation}
remains to be discussed.

The contribution $u^2/c^2\cdot\gamma\cdot(p_0dV_0)$ is a work term,
coming solely from the Lorentz contraction of the volume element,
$dV_0$, in the rest system, when observed in the laboratory frame
(cf. formula (\ref{27})). Such a contribution occurs also in the
energy transformation law and a similar one plays the role of explicit
pressure work, $-pdV$. We do not see, that an additional force is
involved with this pure contraction effect. So we decide to delete
this term. There remains the term 
\begin{equation}u^2/c^2\cdot\gamma\cdot
  (dp_0V_0)=u^2/c^2\cdot\gamma^2\cdot (dpV)       \end{equation}
This term occurs also in the total variation of the (internal) energy
of a moving system.

A changing pressure means also a changing applied force (in addition
to a change in internal energy, $dE$) as equilibrium has to be
maintained in quasi-static processes. So it seems reasonable to
associate the term really with an applied \tit{moving force}. So we
finally conclude
\begin{statement} In relativistic thermodynamics the only work terms
  we are taking into account are i) work of pressure, $-pdV$, ii)
  adiabatic translatory work, $\mbf{u}\,d\mbf{G}$ but with
  $E_0=m_0c^2= const$, iii) no work is involved in the exchange of
  heat between comoving systems, but we include a work term coming
  from $dp_0V_0$. I.e. we assume
\begin{equation}\delta W=-pdV+u^2/c^2\cdot\gamma^2\cdot (dpV)+E_0/c^2\cdot d(\gamma\mbf{u})
\end{equation}
\end{statement}
\subsection{The Relativistic Concept of Heat}
Heat is a subtle concept even in in non-relativistic phenomenological
thermodynamics. The most straightforward way of introducing it is, in
our view, to regard it as the stochastic, disordered and non-coherent
contribution in the energy conservation law (in contrast to e.g. the
highly organized form of work, which is, in effect, some avaraged and
integrated form of the individual effects of many microscopic
events). So it is reasonable to define it simply by
\begin{equation}\delta Q=dE-\delta W     \end{equation}
that is, as the difference between the increase of internal energy and
applied work. 

From our above line of observations and arguments, its functional
dependence on the proper or rest variables can now be inferred from
the expressions for $E$ and $\delta W$. In formula (\ref{28}) we have
got
\begin{equation}E=(E_0+p_0V_0\cdot u^2/c^2)\cdot\gamma
\end{equation}
Furthermore we have
\begin{equation}p\cdot dV=(p_0\cdot dV_0)\cdot\gamma^{-1}      \end{equation}

If the full $\mbf{u}d\mbf{G}$, is included in the work term (as it is
done in the classical papers), one can proceed as follows:
\begin{equation}dE=(dE_0+d(p_0V_0)\cdot u^2/c^2)\cdot\gamma     \end{equation}
\begin{equation}\delta W=-p_0dV_0\cdot\gamma^{-1}+u^2/c^2\cdot(dE_0+d(p_0V_0)\cdot\gamma    \end{equation}
that is, we can keep the differential, $d$, outside of the product
$(p_0V_0)$, as the respective terms cancel each other. We then arrive
at (see e.g. \cite{Pauli} or \cite{Tolman}):
\begin{equation}\delta Q=\delta Q_0\cdot\gamma^{-1}=\delta Q_0\cdot(1-u^2/c^2)^{1/2}   \end{equation}

We learned in the preceding subsection that some of these contribution
in $\delta W$ are presumably inexistent. We have
\begin{equation}d(p_0V_0)=dp_0V_0+p_0V_0      \end{equation}
The ($dp_0V_0$)-contribution in the variation of the energy is
compensated by the corresponding term in $\delta W$ and we end up with
the formula
\begin{multline}\delta Q= (dE_0+p_0\cdot dV_0\cdot
  u^2/c^2)\cdot\gamma+p_0\cdot dV_0\cdot\gamma^{-1}\\ 
=\gamma\cdot (dE_0+p_0\cdot dV_0(u^2/c^2+\gamma^{-2}))=\gamma\cdot
(dE_0+p_0\cdot dV_0)
   \end{multline}
We thus get
\begin{statement}The infinitesimal element of heat transforms under a
  Lorentz transformation like
\begin{equation}\delta Q=\delta Q_0\cdot\gamma=\delta
  Q_0\cdot(1-u^2/c^2)^{-1/2}   
 \end{equation}
That is, in contrast to energy and work, it transforms as the zero
component of a 4-vector. A related observation was made in
\cite{Moeller} after a long and involved calculation!
\end{statement}

We want to add two remarks. First, we see that a very subtle
compensation has to happen between contributions which transform very
differently under a Lorentz transformation, in order that a coherent
transformation behavior of central quantities like e.g. the heat does
occur. I think, this is one of the main difficulties in this business
if one is really willing to treat the problems in full generality.
Second, we think, the Lorentz covariance of the heat (in contrast to
energy and work) is due to a subtle effect. Work is somehow the
summation over microscopic transfers of energy. Neither (moving) walls
or other external mechanical processes are really involved. So we
think, that each of these elementary (statistical) contributions
transforms as the zero component of an energy-momentum vector. The
same does then hold for the sum of such contributions. This is
completely different for macroscopic (internal) energy or work.
\subsection{The Relativistic Concept of Temperature and the
  Relativistic Carnot Cycle}
The notion of temperature is presumably the most problematical and
subtle one in relativistic thermodynamics and there exists a wide
range of different opinions. We first mention some frequent errors as
to this notion. In the older literature the opinion is sometimes
suggested that an observer in the laboratory system really observes a
higher or lower temperature in a moving body compared to its rest
temperature by somehow reading off a comoving thermometer. This
opinion is almost certainly incorrect as what he will observe is
simply the rest temperature. As a consequence, many of the newer
thought experiments, trying to show that temperature is actually an
\tit{invariant}, are somewhat beside the point (see the section about
the so-called zeroth law of thermodynamics).

To give an example, fixing a certain definite point on the temperature
scale, e.g. the coexistence point of ice and water in the rest system,
nothing spectacular will happen if the system is set into motion in a
quasi-stationary process. That is, ice will not start to melt or
water to freeze. What an observer sees is simply the rest temperature
even if the system moves. This is not some kind of \tit{moving
  temperature}. 

We have in fact to remember that in thermodynamics the \tit{absolute
  temperature} is playing the fundamental role in the thermodynamic
relations. Absolute temperature, on the other hand, is introduced and
defined via the \tit{Carnot process} or plays the role of an
\tit{integrating factor} in the relation between entropy and
heat. Therefore, to begin with, we should concentrate on this notion
and its generalisation.

The corresponding structural relation now fixes the transformation
properties of absolute temperature. From $dS=\delta Q/T$ for a
reversible process and $dS=dS_0$, $\delta Q=\delta Q_0\cdot\gamma$, it
follows
\begin{ob}T transforms under a Lorentz transformation as
\begin{equation}T=T_0\cdot\gamma     \end{equation}
\end{ob}
Furthermore, as e.g. described in \cite{Becker}, the Carnot cycle
allows us to define and measure absolute temperature via the universal
relation
\begin{equation}\eta=(Q_1-Q_2)/Q_1=:(T_1-T_2)/T_1    \end{equation}
and $\eta$ the \tit{Carnot efficiency}. On the other hand, heat can be
measured independently of the concept of temperature as is beautifully
described in e.g. \cite{Callen}, chapt.1.7, via purely mechanical processes.   

In order to show that all this is not a purely theoretical construct,
one can analyse the \tit{relativistic Carnot cycle}, as discussed in
different realisations in e.g. \cite{Laue} or \cite{Tolman}.\\[0.3cm]
Remark: Note that both Laue and Tolman belong to the classical
period. I.e., the temperature of the moving system is lower than the
rest temperature. As a consequence, some of the occurring plus or
minus signs are different from our treatment below.

Consider now a simple system (the engine), working at constant
pressure, $p=p_0$ (i.e. terms, containing a $dp_0V_0$, do not!
contribute), over the whole reversible cycle and operating between a
reservoir, $R_1$, being at rest in the laboratory frame, having
temperature $T_1$, and a reservoir, $R_2$, moving with the velocity
$u$, and having the temperature $T_2=T_1\cdot\gamma$. The system may
initially be at rest in a state, described by energy $E_a$, volume
$V_a$ and temperature $T_1$. Let it now absorb the amount of heat
$Q_1$ from $R_1$ at constant pressure and doing the work $p\cdot
(V_b-V_a)$. We then have
\begin{equation}Q_1=(E_b-E_a)+p\cdot(V_b-V_a)=H_b-H_a    \end{equation}
(cf. observation \ref{enthalpy}).

The second step consists of a reversible adiabatic acceleration to the
velocity $u$ of reservoir $R_2$, with the internal conditions
unaltered. The work done by the system is
\begin{equation}W_2=E_b-E_c   \end{equation}
with $E_c$ given by e.g. formula (\ref{25}). In the third step the
amount of heat being released to the reservoir $R_2$ is $Q_2$ and the
amount of work done is 
\begin{equation}W_3=p\cdot (V_d-V_c) \end{equation} We assume now that
the amount of released heat in the third step is just sufficient so
that the system can be returned to its initial state by a reversible
deceleration. This is exactly the case (see below) if
\begin{equation}Q_{2,0}=(E_d-E_c)_0+p(V_d-V_c)_0=-Q_1=-Q_{1,0}    \end{equation}
holds, with the subscript $0$ denoting the respective proper values of
the quantities, i.e. for $Q_{2,0}$ it is the amount of heat measured
by a comoving observer. The work done by the system is
\begin{equation}W_4=E_d-E_a   \end{equation}

The first law of thermodynamics tells us that
\begin{equation}Q_1+Q_2=W_1+W_2+W_3+W_4  \end{equation}
with $Q_i$ the heat absorbed by the system and $W_i$ the work done
by the system, or
\begin{multline}Q_2=(pV_d+E_d)-(pV_c+E_c)=H_d-H_c=\\
(pV_a+E_a)\cdot\gamma-(pV_b+E_b)\cdot\gamma=(H_a-H_b)\cdot\gamma =\\
-(p(V_b-V_a)+(E_b-E_a))\cdot\gamma=-Q_1\cdot\gamma  \end{multline}
(according to formula (\ref{28})). We see that in each cycle there is
a heat transport from reservoir $R_1$ to reservoir $R_2$ and a
negative amount of work done by the system on the environment, i.e.
\begin{equation}\Delta W=Q_1\cdot(1-\gamma)<0     \end{equation}

It is clear that by reversing the direction of the process we can
extract a positive work out of the compound system (i.e., system plus
reservoirs) while now heat is transported from $R_2$ to $R_1$.
\begin{ob}It is important to note that nowwhere in the calculations it
  was really used that the reservoir $R_2$ has a higher
  temperature. Only the first law of thermodynamics was
  exploited. What we however implicitly assumed is that system and
  reservoirs coexist thermodynamically at relative velocity zero, so
  that the ordinary laws of thermodynamics can be applied
  (e.g. quasi-static heat exchange). 
\end{ob}

The second law now tells us that
\begin{equation}\label{63}Q_2/Q_1=-T_2/T_1=-\gamma    \end{equation}
i.e.
\begin{equation}T_2=T_1\cdot\gamma    \end{equation}
That is, the relativistic Carnot cycle allows us to give an
operationalistic definition of absolute temperature as in ordinary
thermodynamics and exhibits the internal consistency of the
framework.\\[0.3cm]
Remark: In the classical framework, i.e. with a work term coming from
the influx of heat, the work done by the system is positive if heat is
transported from $R_1$ to $R_2$, and vice versa.\vspace{0.3cm}

The Carnot efficiency in our reversed cycle is
\begin{equation}\eta=1-Q_1/Q_2=1-T_1/T_2=1-\gamma^{-1}=1-\sqrt{1-u^2/c^2}>0     \end{equation}
We hence can conclude that our reversed Carnot process does work on
the environment in an objective sense, the reason being mainly the
Einstein-equivalence of matter and energy and not! so much a
particular assumption about \tit{moving temperature}. The energy or
mass which is decelerated is larger than the mass which is accelerated
due to the additional absorption of heat.\\[0.3cm]
Remark: It is important to note that the above Carnot cycle is of a
very special type. This will be discussed in connection with the
so-called zeroth law of thermodynamics, see below.

\subsection{The Covariant Expressions of Heat and Temperature}
We have seen from our preceding discussions that heat and temperature
transform as
\begin{equation}\delta Q= \delta Q_0\cdot\gamma\quad , \quad T=T_0\cdot\gamma     \end{equation}
if the initial system is the rest system. For general moving systems,
moving with the velocities $\pm \mbf{U}$ relative to each other, we
thus have to conclude that the transformation laws are those of
4-vectors.

It is straightforward to build 4-vectors from scalars in the rest
frame. For temperature and heat the covariant generalisations are
hence
\begin{equation}T^{\mu}=(T_0\cdot\gamma,T_0\cdot\mbf{u}/c\cdot\gamma)\quad
, \quad \delta Q^{\mu}=(\delta Q_0\cdot\gamma,\delta Q_0\cdot\mbf{u}/c\cdot\gamma)\end{equation}
i.e.
\begin{equation}T^{\mu}=T_0/c\cdot U^{\mu}\quad , \quad \delta
  Q^{\mu}=\delta Q_0/c\cdot U^{\mu} \end{equation}
with $U^{\mu}$ the 4-velocity.

The covariant generalisation of the second law of thermodynamics is
then
\begin{equation}dS\geq \beta^{\mu}\delta Q_{\mu}=\beta^0\delta Q^0-\mbf{\beta}\cdot\delta\mbf{Q}     \end{equation}
A certain warning should be spelled out. In the rest system we have of
course that $\beta^0(0)=(T^0(0))^{-1}$, with the velocity of the
system relative to the respective IF given in braces. In a IF moving
with the velocity $\mbf{u}$, we however have 
\begin{equation}\beta^0(u)=\gamma\cdot\beta^0(0)=\gamma\cdot(T^0(0))^{-1}=\gamma^2\cdot T^0(u)^{-1}  
\end{equation}
\begin{ob}In a moving IF the relation between $\beta^0$ and $T^0$ is
  different from the standard form, we have in the rest system. 
\end{ob}
Remark: See in this connection e.g. formulas (4.19),(4.20) in
\cite{Ojima}, where it is concluded that $T^0(u)^{-1}=\beta^0(u)$,
which would lead to $T=\gamma^{-1}\cdot T_0$ instead of $T=\gamma\cdot
T_0$. We have however seen that both $\beta^{\mu}$ and $T^{\mu}$
transform as 4-vectors.
\section{\label{zeroth}The Covariant Expression of the Zeroth Law of Thermodynamics}
A better understanding of the relativistic zeroth law (which in this
context we take to be a statement about the respective temperatures of
thermally coexisting subsystems) is particularly important, as it is,
so to speak one of the pillars of ordinary therrmodynamics.
Furthermore, its seeming violation was invoked by e.g.  Landsberg (as
was already mentioned in the introduction; \cite{Landsberg3}, p.334)
to construct a paradoxon in case one is willing to assume that
temperature is not! a scalar. We will now show that this conclusion is
ill-founded and that, perhaps coming as a surprise, the zeroth law
looks different in relativistic thermodynamics.

More specifically, we show that the transformation law,
$T=T_0\cdot\gamma$, we are favoring in this paper, is not! in
contradiction with an appropriately formulated zeroth law.
\begin{statement}(Entropy-maximum principle) An isolated system being
  at rest (i.e. having constant rest energy) and possibly consisting
  of several subsystems which can e.g. exchange energy with each
  other, is in thermal equilibrium if its entropy is maximal.
\end{statement}
In ordinary thermodynamics equally of the temperatures, $T_1$, $T_2$, \ldots,
of the subsystems can immediately be derived from this principle.

The discussion of the relativistic case is more subtle and we proceed
as follows. We employ the results of the discussion of the
relativistic Carnot cycle in the preceding section. Note that the
cycle is very special in several respects if compared with the
ordinary Carnot cycle. We found that both
\begin{equation}Q_{2,0}=-Q_{1,0} \quad , \quad Q_2=-\gamma\cdot
  Q_1=-\gamma\cdot Q_{1,0}    \end{equation}
and
\begin{equation}T_2=T_1\cdot\gamma=T_{1,0}\cdot\gamma    \end{equation}
have to hold. Such strong constraints do not exist in the ordinary
Carnot cycle.

We can now use this relativistic Carnot cycle in the discussion of the
zeroth law. We treat the reservoirs as large but finite subsystems (as
compared to the engine, which can be chosen infinitesimal if
necessary). The compound system is assumed to be closed in the sense
of formula (\ref{73}). The engine is
now used to reversibly transport a small amount of heat or internal
energy from the one system to the other under the proviso
\begin{equation}\label{73}E_{1,0}+E_{2,0}= const    \end{equation}
This is exactly the kind of variation which is used in the ordinary
derivation of the zeroth law, where now $E_i$ are the internal
energies of the (finite) reservoirs (or subsystems) while we now write
$\Delta Q_i$ or $\Delta Q_{i.0}$ for the exchanged amounts of heat
with
\begin{equation}\Delta Q_{2.0}=- \Delta Q_{1.0} \quad , \quad \Delta
  Q_2=-\gamma\cdot \Delta Q_1
    \end{equation}

    We have thermal coexistence of the two subsystems if such an
    infinitesimal exchange of heat does not change the total entropy.
    Note in this respect that quasi stationary acceleration or
    deceleration of the engine does not change the entropy. By
    definition, the entropy (as a state function) of the engine does
    not change in a complete cycle. This implies
\begin{equation}Q_2/T_2+Q_1/T_1=0    \end{equation}
(cf. formula (\ref{63})). But exactly the same formula holds for the
two reservoirs (subsystems). I.e., we have
\begin{equation}\Delta S_2+\Delta S_1=0   \end{equation}
for the total variation of the entropy of the compound system.
\begin{conclusion}The two subsystems (of the closed compound system)
  are in thermal equilibrium if
  \begin{equation}T_{2,0}\cdot \gamma=T_2=T_1\cdot\gamma=
    T_{1,0}\cdot\gamma \end{equation} that is, if the rest temperature
  of the moving system is identical to the rest
  temperature of the other system. But observed from the laboratory
  frame, the temperatures differ in just the above sense.
\end{conclusion}
Remark: The only place where we found this point discussed in a
however qualitative manner is in \cite{Laue}. V.Laue provides also another
thought experiment which is equally convincing (see p.178).
\section{\label{Ott}Some Remarks on a Paper by Ott}
As we already remarked in the introduction, we think that also the
analysis by Ott is not really complete and contains gaps. The reason
is however understandable as some of the problems are in fact
well-hidden. Typically (as is the case in quite a few other
investigations), it turns out that the processes being analysed are
too restricted and special. This holds in particular for his version
of Carnot-cycle (cf. section 2 of \cite{Ott}).

For one, as usual the pressure is kept constant. By making this
assumption he avoided to discuss the critical term, $dp_0V_0$, we
discussed in our analysis of the first law of thermodynamics and which
is both a contribution in the problematical work term,
$\mbf{u}d\mbf{G}$, and the internal energy. For another (and strangely
enough), the volume is not! changed in his Carnot cycle during the
step in which the engine absorbs heat from a reservoir. He says that
the occurring work is somehow stored within the system? This is in our
view very funny and difficult to understand.

A thermodynamical state describing the coexistence of e.g. fluid and
steam allows for moves on the state manifold in which the pressure
remains constant but then the volume changes if fluid is transformed
into steam or vice versa. Otherwise the temperature will change, but
in this part of the cycle the engine is in contact with a heat
reservoir! Therefore the only pressure work in Ott's Carnot cycle
occurs in the adiabatic parts where the engine is accelerated or
decelerated. But in our view it is inconsistent to have a state change
with all! state variables kept fixed.

For all these reasons Ott gets an extremely simple (almost trivial)
version of the first law (cf. his formulas (13), (16) and (17)) which
reads
\begin{equation}\delta Q=dE=dE_0\cdot\gamma \label{first}    \end{equation}
In short, Ott gets the same transformation laws as we did but only for
a very incomplete situation.

As we have argued above, the crucial problem is to deal with terms
like
\begin{equation}pdV=p_0dV_0\cdot\gamma^{-1}      \end{equation}
with $dV_0\neq 0$ in the rest system in view of the completely
different transformation behavior of e.g.
\begin{equation}d(E+pV)=d(E_0+p_0V_0)\cdot\gamma      \end{equation}
This led to a deeper analysis of the nature of work in relativistic
thermodynamics. 
This problem is completely lost sight of if one deals only with very
simplified processes (the same remark applies to the approach of
Rohrlich being dealt with in the following section).

A further problematic point can be found on p.76 in formula (9). Ott
obviously takes for granted that the thermodynamic energy is the zero
componemt of a 4-vector. We think it is common knowledge that for
non-closed systems as in relativistic thermodynamics this is not! the
case (cf. our preceding sections). Consequently Ott concludes from his
simplified (but presumably wrong!) version of the first law (formula
(\ref{first})) that $\delta Q$ is also the zero component of a
4-vector. 

This however follows in a complete analysis only from a series of not
entirely trivial steps (cf. the preceding sections or the analysis
given by Moeller in his book \cite{Moeller}). This assumption, made
by Ott was also criticized by Balescu in \cite{Balescu}. As far as we
can see, Arzeli\'{e}s makes a similar assumption. This is only correct
if parts of the exterior are included in the system, which is in our
view however not desirable.

\section{Some Remarks on a Paper by Rohrlich}
We also want to briefly comment on an approach, developed by Rohrlich
(\cite{Rohrlich}). The general tenor is that there are \tit{true}
transformation properties and, on the other hand, only \tit{apparent}
ones in special relativity. We must confess that we do not appreciate
very much this distinction. For example, the Lorentz contraction of
length or volume is only an apparent one in the philosophy of
Rohrlich and he rather suggests a volume transformation which is
4-vector like. 

It is not the place here, to dwell in more detail on the philosophy,
underlying Rohrlich's paper. While it is possibly shared by other
physicists, we would like to make our own point of view clear, namely
we regard these phenomena as real (as also e.g. Rindler does in his
book \cite{Rindler}). A nice thought experiment as to this question is
developed by Bell in his essay about special relativity (\cite{Bell}),
which clearly shows that Lorentz contraction is real.

There is however in section 3 of Rohrlich's paper a treatment of the
transformation law of relativistic temperature which is a little bit
different from the ordinary one and arrives at a result which at
first glance differs from our own findings. We think this point is a
good example to learn how tricky calculations in this field actually are.

Rohrlich writes the first law in the rest system slightly differently
as
\begin{equation}T_0dS=dH_0-V_0dp_0    \end{equation}
with again, $S$, $p$, treated as invariants and $H_0=E_0+p_0V_0$. He
than makes the seemingly innocent assumption that the first law in
this form holds also in an arbitrary IF. I.e., he assumes
\begin{equation}TdS=dH-Vdp           \end{equation}  
and concludes from this that 
\begin{equation}T=T_0\cdot\gamma^{-1} \end{equation} must hold, where
he treated only the special case, $dH=dH_0=0$ anyhow. I.e., he gets
the classical transformation law. The inconsistency of his assumption
possibly escaped Rohrlich because he really treated only the case
$TdS=-Vdp$.

We showed however above that $H$ transforms as the zero-component of a
4-vector, i.e. we actually have
\begin{equation}H=H_0\cdot\gamma \end{equation} Furthermore, we showed
that the complete work term is in fact more complicated and consists
of more than only pressure work. We in fact get for the general case
\begin{equation}\gamma^{-1}\cdot TdS=T_0dS=dH_0-V_0dp=\gamma^{-1}\cdot
dH-\gamma\cdot Vdp\end{equation}
hence
\begin{equation}TdS=dH-\gamma^2\cdot Vdp      \end{equation}
that is
\begin{conclusion}The first law of thermodynamics in the above form,
  given by Rohrlich, does not transform in a Lorentz invariant
  way. Therefore the classical transformation result of temperature
  does not follow either.
\end{conclusion}
\section{Commentary}
We hope that it has become clear from our discussion that various of
the papers existing in the field contain gaps or arguments which are
not conclusive. We also emphasized that the wide spread habit to
discuss only very particular and restricted thermodynamic processes is
dangerous and leads sometimes to wrong conclusions concerning the
correct transformation laws of thermodynamic variables. 

This holds in particular for the various contributions entering in the
(infamous) work term $\mbf{u}\cdot\mbf{G}$. We showed that some of
them are in fact zero but there is one contribution which survives.
This term has been overlooked in the past because most of the authors
treated only processes at constant pressure.

Furthermore, a detailed analysis of the so-called zeroth law showed
that, in contrast to what some groups claimed, thermodynamic systems,
moving with a non-vanishing relative velocity, $u$, can thermally
coexist, provided their (moving) temperatures fulfill
\begin{equation}T_2=T_1\cdot\gamma(\mbf{u})     \end{equation}
which however implies that their respective rest temperatures are
equal, i.e. $T_{2,0}=T_{1,0}$.

It should again be emphasized that our guiding principle was that the
first and second law of relativistic thermodynamics remain form
invariant with respect to the transformed variables. While heat turns
out to be a 4-vector, this was not the case for (internal) energy and work.

\end{document}